\documentclass[aps,prl,preprint,groupedaddress]{revtex4}

\begin{document}


\title{S. Chandrasekhar: White Dwarfs, $H^-$ ion,.., Black holes, Gravitational waves}


\author{Patrick Das Gupta}
\affiliation{Department of Physics and Astrophysics, University of Delhi, Delhi - 110 007 (India)}
\email[]{patrick@srb.org.in}



\begin{abstract}
This is a concise review of S. Chandrasekhar's research contributions to astrophysics, ranging from his early studies on white dwarfs using  relativistic
 quantum statistics  to topics
 as diverse as  dynamical friction, 
negative hydrogen ion, fluid dynamical instabilities, black holes
 and gravitational waves. The exposition is based on simple physical explanations
 in the context of observational astronomy, addressed primarily to the undergraduate students.
 Black holes and their role as central engines of active, compact, high energy sources have been discussed in some details.

\end{abstract}


\maketitle

\section{I Introduction}

The impactful research journey of  Subrahmanyan Chandrasekhar began on  July 31, 1930,
  from Bombay port on a ship. The 19 year old
   Chandra was on his way to England for higher studies.  Armed with his understanding of  Fowler$^\prime$s work  on white dwarfs$^1$, Chandra was 
  immersed in the mathematical equations describing these dense objects, during that voyage. He had realized that Fowler's theory needed modification, since for sufficiently
 massive white dwarfs, particle number densities could be so high that a large fraction of electrons would be occupying  very high 
energy levels, moving with relativistic velocities. 
 
 At this juncture, a quick summary of  stellar evolution
  theory is in store. In main sequence  stars (like  Sun), nuclear fusion of hydrogen to helium 
   supplies the  required
  thermal energy  to stall gravitational contraction of a star, enabling  it
   to attain a quasi-hydrostatic
  equilibrium.
  As the star advances in age, a further sequence of nuclear fusion reactions gets activated in its
  core - helium burning to carbon and 
 oxygen, carbon burning to sodium and magnesium and so on, if the star is massive enough, till  the formation of  iron-rich core.
  Iron nucleus being the most stable one, further nuclear burning is energetically not feasible. As the iron core cools, the thermal pressure due to the left over
 heat energy is 
 insufficient to halt the collapse under its own weight. Eventually  the electron density becomes so high that the electron degeneracy
  pressure comes into play, and prevents further contraction. For low mass stars however, white dwarf stage is reached with the formation of a degenerate carbon-oxygen core
  after the red giant phase resulting from helium burning.
 
  Degeneracy pressure is a consequence of quantum statistics in extremely dense matter. Pauli exclusion principle (PEP) states that no two identical fermions can have the 
 same state. Electrons, protons, neutrons, neutrinos, etc., being spin half particles, are fermions. According to PEP, in a
  gravitationally bound system like the iron-rich core of an evolved  star,  all the electrons cannot occupy
  the lowest energy level (unlike, what happens to identical bosons in Bose-Einstein condensates, e.g. He-4 superfluid). So, the energy levels are filled up with two electrons
  (one with spin up state and the other with spin down) per orbital, as demanded by the PEP. More the density of electrons, higher is the electronic energy level that gets to be 
  occupied.  
 
 Gravitational shrinking of such a dense core  leads 
 to an increase in electron density, thereby encountering a stronger resistance since the contraction implies  putting 
  electrons at higher energy levels. Therefore, in such a `degenerate matter' dominated system, the gravitational collapse instead of lowering 
 the total energy of the star tends to increase it. The ensuing pressure against shrinking, 
 arising out of PEP in such an electron-rich dense
  matter is called electron degeneracy pressure (EDP).  A white dwarf is a star that is in hydrostatic
  equilibrium not because of thermal pressure but  due to the EDP  that counteracts gravitational
  contraction. Fowler had assumed electrons to be moving non-relativistically inside the degenerate core, and had arrived at the result that
 the EDP  of a white dwarf is proportional to $\rho^{5/3}$, where $\rho $ is the density of the core$^1$.
 
\section{II Chandrasekhar limit and compact objects}

 Having realized that electrons would be randomly whizzing around in the enormously dense core with relativistic speeds,
  Chandra incorporated Einstein's special relativity in the  analysis of white dwarfs,
 and found that the EDP is actually proportional to $\rho^{4/3}$ instead, demonstrating that the relativistic degeneracy
 pressure does not increase as rapidly as in Fowler's case.   
  Employing a physically realistic analysis of the relativistic problem of a 
dense  star ruled by a polytropic equation of state, in which gravity is countered by the
 EDP, he arrived at
the celebrated Chandrasekhar mass limit$^2$,
 $$ M_{Ch}= \frac {0.2} {(m_p \mu_e)^2} \bigg (\frac {\hbar c} {G} \bigg )^{3/2} \ \  ,\eqno(1)$$
where $\hbar$, $G$, $c$, $m_p$ and $\mu_e $ are the reduced Planck$^\prime $s constant, Newton$^\prime $s gravitational
constant, speed of light, mass of a proton and mean molecular weight per electron, respectively. It is
remarkable that such a significant result concerning stars should be expressible in terms of fundamental 
quantities (except for $\mu_e$). In white dwarfs, the value of $\mu_e $ is about 2, so that from eq.(1) one finds
 the limit to  be $ M_{Ch}\approx
 1.4
\ M_\odot $, where $  M_\odot = 2 \times 10^{30} $ kg is the Sun$^\prime $s mass.

 During the voyage, Chandra was unaware that Anderson in 1929 and Stoner in 1930 had independently
 applied special relativity
to obtain mass limits for a degenerate, dense star of uniform density but without taking into account the 
condition of hydrostatic equilibrium$^{3,4,7}$. When Chandra showed his work to Fowler after reaching 
Cambridge, Fowler pointed out  contributions of Anderson and Stoner to him. Chandra added these references to his papers on relativistic degeneracy 
in white dwarf stars $^5$. Incidentally, Landau too 
had arrived at a mass limit independently in 1931, which appeared in print one year later$^{6}$.

 Chandrasekhar mass limit entails that no white dwarf with mass greater than this limit can hold out
against gravitational collapse. All the white dwarfs discovered so far (e.g. Sirius B, the companion
 star to Sirius), have mass less than $M_{Ch}$.  For masses beyond this limit, two prescient ideas were put forward independently, that 
 played important roles later - one of Landau$^{6}$, before the discovery of neutrons by Chadwick in 1932 and the other 
 by Baade and Zwicky$^{8,9}$, immediately after the discovery.  Landau had speculated that for stellar cores whose mass exceeded $M_{Ch}$, the density  would become so large due to
 shrinking that
 the atomic nuclei in the core would come in contact with each other - the whole core turning into a giant nucleus$^{6}$. Baade and Zwicky, while attributing the origin of cosmic rays to 
 stellar  explosions called supernovae, correctly identified the energy liberated due to sudden decrease in the gravitational potential energy (as the core collapses rapidly to
 form a neutron star of radius $\sim 10$  km) as the one that powers supernova explosion$^{8,9}$. A core with mass $M_c$, shrinking from a large size to a neutron-rich ball with radius $R_c$, 
has to give up an energy,
 
 $$E_{exp}\sim \frac {G M^2_c} {R_c} \ \ ,\eqno(2)$$

 since its gravitational potential energy  decreases to $\sim - E_{exp} $. For a 1.4 $M_\odot $ core collapsing to form a neutron star of radius $R_c \approx $ 10 km,
the energy $E_{exp}$ available for explosion  is as high as $\sim 10^{53}$ ergs.

 Why does the core become neutron-rich? As the core shrinks, its density rises till it reaches nucleonic values
  $\sim 10^{12}$ - $10^{14}\  gm/cm^3$, when protons in the core transform into neutrons by capturing electrons and 
 emitting neutrinos$^{10}$. Neutrinos, being weakly interacting particles, escape from the core. While in the neutron-rich core, the neutron degeneracy pressure (arising from PEP,
 as neutrons too are
spin half particles) prevents further gravitational contraction, resulting in the formation of a neutron star.

 With the detection of periodic emission of radio-pulses from a
source  by Jocelyn Bell and Anthony Hewish in 1967, existence of neutron stars as pulsars was established. 
 Pulsars are rapidly spinning neutron stars with rotation period ranging from about few milli-seconds to few seconds. The observed pulses are due to electromagnetic radiation from
 accelerated charge particles moving along strong magnetic field lines inclined with respect to the rotation axis (The polar magnetic field strengths vary from $\sim 10^{10}$ to
 $\sim 10^{14}$ gauss).
Recently, a milli-second pulsar was found to have a mass of $\approx 2 \ M_{\odot} $, determined using a general relativistic effect called Shapiro delay in which
radiation grazing past a compact, massive object, arrives at the observer with a time lag because of the strongly curved space-time geometry it encounters near the massive
 star$^{11} $.

 As long as the core is lighter than about 
$2 - 3 \ M_{\odot}$, it can survive as a neutron star (The 
 mass limit in this case is uncertain as it depends crucially on the equation of state for nuclear matter which, for such huge densities existing inside neutron stars,
 is unknown$^{11, 12}$).  
  The released neutrinos, after travelling long distances, eventually lose their energy to the stellar envelope, causing the latter to be blown apart, giving rise to a Type II supernova.
  Measurements concerning detected neutrinos from the supernova SN 1987A indicate that these ultra-light, weakly interacting particles carry away $99 \% $ of the gravitational binding energy released
from the collapsing core, lending credence to the neutrino driven explosion models$^{10}$.
 
   The observed masses for neutron stars do not appear to exceed $\sim 3 \ M_{\odot} $$^{11,12} $, suggesting that a massive star whose core is heaver than this limit, would certainly 
 collapse to
form a black hole. The long duration gamma ray burst sources that exhibit prompt gamma emissions with photons having energy predominantly in 0.1 - 1 MeV range, 
and lasting for about 2 - 1000 s are likely to be massive cores collapsing to form black holes$^{13}$.  
Eddington was extremely critical of Chandra's work which suggested that a
 star exceeding the mass limit would shrink gravitationally to a point singularity$^{14} $. Three decades later, Penrose and Hawking, using Raychaudhuri equation, proved the seminal 
singularity theorems. According to these theorems,  gravitational collapse of normal matter generically lead to formation of singularities,
namely, creation of black holes$^{15-17} $. 

\section{III Dynamical Friction}
 Chandra played a stellar role in the research area of gravitational dynamics of point masses, from 
 1939 to 1944, 
 that culminated in his celebrated papers on dynamical friction$^{18,19}$. 
 There is abundance of 
 gravitationally bound systems of massive objects in the cosmos like globular clusters, galaxies, clusters
 of galaxies, etc. Entities that make up these bound systems, apart from moving in self-created
  gravitational potential wells, also suffer two-body gravitational encounters, resulting in
  exchange of energy and momentum. Chandra discovered that a
 massive body in motion, surrounded by a swarm of  other less massive objects, suffers
  deceleration that is proportional to its mass$^{18} $. 
 
  This dynamical friction originates from 
   cumulative gravitational encounters that the massive body experiences due to the presence 
 of other objects in the background. The physical cause of dynamical friction  can be 
 intuitively understood by going to the reference frame in which the more massive body is at rest.
  In this reference frame, the swarm of background objects moving past the 
 massive body get focussed  behind it because of its attractive force,  forming a wake of higher number density of objects, and hence of higher mass density. When we switch back to the original frame in which 
 the massive body is moving, we find that the
  mass density of the wake behind the larger mass is greater than the density ahead. Consequently,  it experiences a 
  greater gravitational pull from behind, and thereby suffers a
 gravitational drag force whose magnitude is proportional to the square of its mass and
 inversely proportional to the square of its speed$^{20, 21}$.
 
 Observational support for dynamical friction comes from  sinking of globular 
 clusters towards the central regions of galaxies, and galactic cannibalism 
 in which the orbit of a satellite galaxy decays, leading eventually to its merger with the
 bigger galaxy$^{21,22}$. 

\section{IV Negative Hydrogen Ion}

 During the early forties, Chandra was also  involved with the quantum theory of 
 negative hydrogen ion.  Now, can a proton capture two electrons to form a charged bound 
 state? How is it relevant to astrophysics? The first  issue had been settled by Bethe in 1929 who showed that
 quantum mechanics indeed predicts formation of  
 $ H^- $ ions$^{23} $.  As to the second question, it has been
  found over the years
 that $ H^- $ is a weakly bound system with  a binding energy of $\approx $ 0.75 eV.
  Since it takes only about 0.75 eV to knock off the extra electron from 
 $ H^- $, its life-time under terrestrial conditions is small but  in thin and tenuous plasma where 
 the collision frequency is low, one expects negative  hydrogen ions to survive for longer
  duration. 
 
 Early on,
 Wildt had foreseen that $ H^- $ would form in large numbers in the upper 
 atmosphere of Sun, because of the abundant presence of hydrogen atoms and electrons there. He had also come to the conclusion that 
  photo-detachment of $ H^- $ would contribute greatly to solar opacity, since radiation from Sun would be attenuated as they photo-ionize
 $ H^- $ ions on their way out$^{24-26}$.

  Chandra and his collaborators played a significant role in calculating 
  $H^-$ photo-absorption matrix element, so crucial for estimating the quantum probability (and, therefore, the cross-section) of photo-ionization of negative hydrogen ion$^{27-33}$. 
The opacity or the optical depth is proportional to the photo-absorption cross-section $\sigma $ as
  well as $n$, the number density of $ H^- $. This is because, the number of photo-ionizations per photon per unit time is $c \ n\ \sigma $, so that the  
 mean free path length for photons is simply,
$$l=\frac {1} {n\ \sigma} \ \ .$$
The optical depth  essentially is  the ratio of the geometrical path length traversed by the radiation to mean free path length $l$ (i.e., it is the number 
of absorptions suffered by the photons on an average).

The negative hydrogen ion has 
 only the ground state as a bound state. There is no singly excited state that is a bound state. As a result,  photons with energy above 0.75 eV, executing random walks out of  Sun due to
 multiple scatterings, would be absorbed by $H^-$ ions after detaching their extra electrons  to the continuum. This is the major cause for solar opacity in the infra-red to visible range of
the electromagnetic spectrum. 
 
 In 1943, Chandrasekhar and Krogdahl drew attention to the fact that the dominant 
 contribution to the
  matrix element describing the probability amplitude for such a photo-dissociation  comes from the wavefunction at large distances (several 
 times the Bohr radius). Therefore, an accurate knowledge of electronic  wavefunction
  of  $H^-$ is required to compute the matrix element correctly$^{27}$.

 Chandra and his collaborators made seminal contributions in calculating the 
 continuous absorption coefficient $\kappa_\lambda $ as a function of the
 photon wavelength  $\lambda $, taking into account  
 dipole-length and dipole-velocity formulae, that provided a solid theoretical foundation
  for the characteristic $\kappa_\lambda $ - $\lambda $ plot which exhibits a rise
 in the range 4000   to 9000 angstroms and  then drops to a  minimum at 16000 angstroms, with a 
 subsequent
  rise$^{34}$.
 
  The negatively charged hydrogen  ion also plays an important role in cyclotrons and particle accelerators$^{35}$. The advantages in making use of  $H^-$ arise  out of  the possibility of
  accelerating them by applying electric fields and obtaining hot neutral beams in Tokamaks (like in ITER)$^{36}$. This is because of the relative ease in detaching its extra electron when 
   $H^-$ ion is present in the gas cells.

\section{V Magnetohydrodynamics} 
 Astrophysical entities are usually permeated with magnetic fields, be it planets like earth, Jupiter, etc., Sun, 
  sunspots, stars, flares, spiral arms of Milky Way, galaxies, and so on. Magnetic field in a conducting
 medium like metal or plasma wears out due to Ohmic dissipation.  How does terrestrial 
 magnetic field, presumably generated by the electric currents flowing in the molten, conducting and rotating core of  Earth, counter Ohmic decay? 
 
 Dynamo theories involving differential rotation and convection in conducting fluids are
 invoked to solve this 
  riddle. 
  However, Cowling had proved that magnetohydrodynamical flows with  axisymmetric geometry 
 will always entail a decaying magnetic field$^{37}$. 
  About two decades later, Backus and Chandra  generalized Cowling$^\prime $s theorem$^{38}$.
 In this context, Chandra studied the possibility of lengthening the decay duration so that an axisymmetric dynamo provides a feasible explanation for geomagnetism$^{39}$.
  It was immediately followed by a paper in which  Backus showed
 that the increase was not large enough to be of geophysical interest$^{40}$.
 Chandra studied several fluid dynamical stability problems employing 
 variational
 methods that have interesting consequences$^{41, 42}$.
 
 An evolved  binary system, consisting of a Roche lobe$ ^{20}$ filling star, spewing out gaseous matter, and a 
 massive compact object (MCO) like a neutron star or a black hole (BH), both  going around the common
 centre of mass, very often acts as a luminous source of  
 high energy photons. In such a binary system,  gas leaking out from the bloated star cannot radially 
 fall on the MCO as it has angular momentum.  
 Instead,  it spirals inwards, forming  an accretion  disc around the MCO so that each tiny
  gaseous volume element of the disc moves along a circular Keplerian orbit$^{43}$.

 For a thin disc with a total mass much less than the mass 
 $M$ of the MCO, the Keplerian speed $v(r)$ of a fluid element at a distance $r$ is given by,
  $$v(r)=\sqrt{\frac {G M} {r}}\ \ ,\eqno(3)$$ 
 Eq.(3) implies that the fluid elements  
  of the accretion disc rotate differentially. Farther the element  
   from the MCO,  lower is its circular speed. Differential rotation leads to viscous rubbing of neighbouring fluid elements at varying distances, causing the accretion disc to  heat up.
 If the disc is sufficiently hot, it 
 emits copious amount of electromagnetic radiation with a spectrum ranging from visible wavelengths to UV photons and X-rays.
 
  There are strong observational evidences that the rapidly time varying, 
 intense X-ray sources, like  Cygnus X-1, are  accreting black holes (see section VII).
  Essentially,
 the gravitational potential energy  of the gas spiralling in, 
  gets converted into
  radiative energy at the rate corresponding to a luminosity of,
$$ L = \epsilon \frac {GM \dot{m}} {r_{min}} \ \ ,\eqno(4)$$
 where $\dot {m} $, $r_{min}$ and $\epsilon $  are the rate of mass accretion, minimum distance reached by the infalling gas and an efficiency factor for the conversion of gravitational
energy to radiation, respectively. The importance of accretion on to compact objects is evident from eq.(4), since source luminosity is larger for smaller values of $r_{min}$.
Similarly, a luminous source requires larger rates of accretion and higher conversion efficiencies.

 For the efficiency factor $\epsilon $ to be large, the accretion disc
 is required to have a high viscosity.  The physics of the mechanism responsible for 
 large viscosities in the disc  is an active area of research.
 Interestingly,  as shown by Balbus and 
 Hawley in 1991,  the Chandrasekhar instability might be the key to the origin of accretion disc
 viscosity$^{44}$.
 Chandra had pointed out that a differentially rotating, conducting and
 magnetized incompressible fluid in a cylindrical configuration, is  unstable with respect
 to oscillating axisymmetric perturbations$^{41}$. 
 
 While investigating Rayleigh-Benard convection in conducting and viscous fluids
 threaded with magnetic field,
 Chandra studied the onset of convection and its dependence on a dimensionless
  number $Q$, representing
 the square of the ratio of magnetic force to  viscous force$^{41}$. Today, this number $Q$ is 
 referred to as Chandrasekhar number (or, also as the square of Hartmann number).
 Chandra made several other contributions in the field of plasma physics and
  magnetohydrodynamics that had far reaching consequences$^{45}$.

\section{VI Chandrasekhar-Friedman-Schutz instability}
 While studying self-gravitating and rotating fluid configurations, Chandra showed that a 
 uniformly dense and uniformly rotating incompressible spheroid  is unstable because of 
 non-radial perturbations, causing emission of gravitational radiation$^{46}$. According to Einstein's general relativity, the curvature of space-time geometry
manifests as gravitational force. Gravitational radiations are wave-like perturbations in the space-time geometry 
 that
 propagate with speed of light, general relativity
 being a relativistic theory of gravitation. Gravitional waves are radiated whenever the quadrupole moment of the mass distribution in a source changes with time.   Friedman and 
 Schutz extended Chandra's findings in 1978, and demonstrated the existence of gravitational wave driven instability in
  the general case of rotating and self-gravitating stars made of perfect fluid$^{47}$.
 
 An intuitive way to comprehend this Chandrasekhar-Friedman-Schutz  (CFS) instability
  is to look at a particular perturbation mode in a rotating star  that is retrograde, i.e. moving in the backward sense relative to the fluid element going around. According to general relativity, 
the space-time geometry around a rotating body is such that inertial frames are dragged along the direction of rotation (This has been recently verified by the Gravity Probe B satellite-borne
experiment$^{48}$). The frame dragging, therefore, would make the retrograde mode appear prograde to
  an inertial observer
  far away from the star. Gravitational waves emitted by this mode will carry positive angular momentum (i.e. having the same sense as the angular momentum of the fluid element) as measured 
in the distant inertial frame. Since, the total angular momentum is conserved, gravitational radiation carrying away positive angular momentum from  the mode, would make the retrograde mode go
 around 
more rapidly in the opposite direction,  leading to an instability. 
 
  In 1998, Andersson demonstrated that a class of toroidal perturbations (the so called r-modes) in a 
 rotating star are generically unstable because of the gravitational wave driven CFS 
 instability$^{49}$.
 It was immediately followed by papers arguing that the r-mode instability would put brakes on the rotation of a newly born and rapidly spinning neutron star$^{50, 51}$. 
  Consequently, as the neutron 
star spins down, a substantial  
 amount of its rotational energy is radiated away as gravitational waves, making it a likely 
 candidate for future detection by the laser interferometric gravitational wave detectors, 
 namely, the LIGOs$^{52,53}$. The CFS instability 
  may soon be put to experimental tests.

 \section{VII Black holes and Gravitational waves}

  Chandra called the astrophysical 
 BHs the most perfect macroscopic objects$^{54}$. 
 Macroscopic entities - like chairs, books, computers, etc.  around us, need an astronomically large number of characteristics each for their description. A sugar cube, for instance,  
  would need not only its mass, density, temperature, but also amount 
 and nature of trace compounds present, the manner in which sugar molecules are stacked, porosity, 
 surface granularities, etc, for its proper specification. In contrast, a BH is 
 characterized by  just three physical  quantities - its mass, charge and angular momentum. 

Schwarzschild BHs do not possess charge or angular momentum, while Kerr BHs  rotate but have
 no charge. On the other hand, Reissner-Nordstrom BHs have charge but do not rotate. Kerr-Newman BHs are theoretically the most general ones, as they possess 
  non-zero mass, charge and angular momentum. Astrophysical black holes are all likely to be Kerr BHs since charge of a BH would get neutralized by the capture
of oppositely charged particles present in the cosmic rays and other ambient matter, and since most cosmic objects possess angular momentum.
 Chandra was particularly fascinated by the stationary, axisymmetric vacuum solutions of Einstein equations
 that described  the Kerr BHs.

BHs are characterized by a fictitious spherical surface 
 called
 the event horizon centred around the point singularity created by the collapse of matter. Nothing can escape from regions enclosed within the event horizon. For a Schwarzschild BH of mass
 $M$, the radius of the event horizon is given by the Schwarzschild radius,
$$R_s = \frac {2 G M} {c^2} = 3 \times 10^6 \bigg (\frac {M} {10^6\ M_\odot} \bigg ) \ \mbox {km}\ \ .\eqno(5)$$
 
 But do BHs exist in the real universe? Classical BHs by themselves  do not 
  radiate. Hawking radiation, which is of quantum mechanical origin, from astrophysical BHs, is too miniscule in amount to be of any observational significance$^{55}$. So, how does one
 find BHs in nature? In conventional astronomy, their detection relies on the presence of gas or stars
  in their vicinity and the ensuing stellar or dissipative gas dynamics around an accreting MCO.
   As discussed in section V, if  the MCO  has an accretion disc around it like in galactic X-ray sources, quasars, blazars or radio-galaxies, the swirling and inward spiralling gas gets heated 
 up,  emitting  radio, optical, UV and X-ray photons, often  accompanied by large scale jets$^{56}$. 

 If gas can spiral down to a distance $r_{min}=\alpha \ R_s$ from the central BH, then according to eqs. (4) and (5) the radiation luminosity is given by,
$$ L = \frac {0.5 \epsilon} {\alpha }  \dot{m} c^2 \ \ .\eqno(6)$$
The real parameter $\alpha $ quantifies the proximity to the central BH.
Eq.(6) tells us that accretion taking place close to the event-horizon can convert an appreciable fraction of rest energy $m c^2$ of the inflowing gas. Higher the accretion rate $\dot{m}$, larger
is the luminosity $L$. (Provided that fluid viscosities in the disc are large enough to give rise to higher efficiencies $\epsilon$, as discussed in section V.) 

The central engine for a quasar or a blazar is, in all likelihood, an accreting 
supermassive BH with $M$ lying in the range $10^7 $- $10^9$ $M_\odot$$^{56}$. Invoking 
  eq.(6) with sufficiently large accretion rates for  blazars, one can theoretically explain  high luminosities (at times, exceeding $10^{48} $ erg/s)  observed in these sources.

  Quasars and blazars also exhibit fluctuating X-ray luminosities on time scales of only few hours. One can derive an upper limit for the size of the
central engine from  causality arguments.  If the observed time scale over which the luminosity varies accreciably is $\Delta t $, the size of the source participating in emission
of   photons
cannot be larger than $ c \Delta t $. This is because, firstly,  every part of the entire region must be causally connected to each other and, secondly, special relativity tells us 
that parts of the region can physically communicate with each other (to remain in causal touch) only with speeds $\leq c$. X-ray variability on time scales of an hour corresponds
to a causal size $\leq  10^9 $ km. Now, from eq.(5), a BH of mass $3 \times 10^8 \ \ M_\odot $ has a Schwarzschild radius of about $10^9 $ km. Short time fluctuations and central engines
involving gas dynamics close to the event horizon of BHs, fit together neatly.

 Observational evidence for accreting super-massive BHs comes not only from short time variability of X-ray fluxes  
  but also from the details of the continuum spectra (e.g. presence of the big blue bump in quasar spectra) observed in these active
 sources. Hence, quasars, blazars and powerful radio-galaxies are most probably distant
 galaxies housing  acccreting supermassive BHs with mass in excess of $10^6 \ M_\odot$ in their central regions$^{56}$. 
 
 Similarly, by monitoring stellar dynamics around the central region of Milky Way for 
 decades, one infers that the Galactic nucleus contains a heavy and compact object, most likely
  to be a supermassive BH
 with a mass of about $4 \times 10^6 \ M_\odot$, within a radius of $10^{13} $ km from
 the Galactic Centre$^{57}$. It is interesting to note that the Chandra X-ray observatory (launched on July 23, 1999, and named after S. Chandrasekhar) revealed the  presence of 
 a X-ray source as well as hot gas with high pressure and strong magnetic field in the 
 vicinity of  the Galactic Centre.

 However, these are indirect detections, implying strictly speaking  the presence of  a very compact, massive central
  object. Inference of an astrophysical  BH, although  very likely, relies on theoretical interpretation.  What happens when
 a BH is perturbed by incident gravitational waves or electromagnetic radiation or Dirac waves describing electrons or neutrinos? Does a
  perturbed BH have a signature emission like a `ringing$^\prime $, 
  analogous to the case of a struck bell? To answer such questions,  Chandra 
 devoted himself to studying BH perturbations from 1970s onwards$^{54,60-67} $.
 
 When a BH is perturbed, the curved space-time geometry around the BH will be subjected to metric fluctuations. For sufficiently small perturbations, 
 a linear analysis of the metric fluctuations can be carried out in terms of normal modes
 except that dissipation due to both emission of gravitational waves as well as their absorption
   by the BH  make the mode frequencies  complex, with the
 decay reflected in the imaginary parts.  In the case of a perturbed BH, such
  quasi-normal modes (QNMs) correspond to a characteristic ringing of the ambient space-time geometry that eventually decays due 
  to emission of gravitational waves.

 QNMs were discovered by Vishveshwara$^{58}$ and Press$^{59}$ while studying gravitational wave perturbations of BHs. Chandra and Detweiler suggested for the 
 first time  numerical
methods for calculating the QNM frequencies$^{62}$.   
 Such investigations throw light on methods for direct detection of BHs. For example, matter falling into  a Schwarzschild BH would lead to excitation of QNMs, resulting in emission of 
gravitational waves  with a characteristic frequency that is 
inversely proportional to the
 BH mass.

 One can understand this dependence from simple dimensional analysis. QNMs would involve perturbations of the event horizon characterized by  the Schwarzscild radius $R_s$ (eq.(5)).
So, the oscillation wavelengths would be typically of a size proportional to $R_s$, making the frequencies depend inversely on the BH mass.
  A supermassive BH with mass $10^6 \ M_\odot$ would ring with a frequency of
 about $10^{-2} $ Hz.  Because of seismic noise, LIGOs cannot detect gravitational waves having such low frequencies. Only a space-based gravitational wave detector like
 LISA (Laser Interferometer Space Antenna) can pick up such low frequency signals from supermassive BHs$^{53,70}$.

Chandra developed innovative techniques to study BH perturbations, and showed that
radial and angular variables could be decoupled to obtain separable solutions for Dirac equation in Kerr background, corresponding to a massive particle (like an electron)$^{62}$. 
Using similar techniques, Don Page extended the separation of variables for massive Dirac equation to the  Kerr-Newman case$^{68}$.
 In 1973, Teukolsky  had separated the Dirac equation
 for two component massless neutrinos in the Kerr background$^{69}$. It will be interesting to investigate if Chandra$^\prime $s technique 
can succeed in separating the Dirac equation for massive neutrinos (with flavour mixing and massive right-handed components included) in the Kerr or Kerr-Newman background.

Kerr BHs possess ergosphere, a region surrounding the event-horizon where test particles with negative angular momenta (i.e. with reverse  sense of rotation relative to BH rotation) can have
negative energy (as measured by a distant inertial observer) orbits. Penrose, in 1969, had shown an ingenious way to extract rotation energy of a Kerr BH that involved sending
 an object that breaks up into two in the ergosphere, with one of the parts going into a negative energy trajectory, while the other escaping with an energy greater than the initial energy
 (since energy is conserved)$^{71}$. 

The wave analogue of Penrose process is superradiance wherein impinging scalar, electromagnetic or gravitational waves emerge out with greater energy after scattering off Kerr BHs.    
Zel$^\prime $dovich was the first to show the existence of superradiance in 1970$^{72}$. Chandra and Detweiler undertook  a thorough investigation of scattering of electromagnetic,
 gravitational 
and neutrino waves in the Kerr background, and showed that neutrinos do not exhibit superradiance$^{73}$. Absence of neutrino superradiance is most likely due to PEP$^{73-76}$.

 Exact solutions of two plane gravitational waves colliding with each other were obtained for the first time by
Szekeres$^{77}$ as well as  Khan and Penrose$^{78}$. Their work showed that due to mutual gravitational focusing, the collision leads
to curvature singularity where gravity becomes infinite. Chandra, along with  Valeria Ferrari and Xanthopoulos, showed that the mathematical
 theory of colliding gravitational waves can be cast in the form of mathematical theory of BHs, and that the formation of curvature singularity due to gravitational
focusing is generic$^{79-82}$.

In the later years, Chandra and Valeria Ferrari studied non-radial oscillations of rotating stars taking into account general relativistic effects$^{83-85}$.  
They showed that the oscillations could be 
described in terms of pure metric perturbations, reducing the problem to scattering of gravitational waves in curved space-time geometry. For strongly gravitating objects like neutron stars,
  such gravitational waves may get trapped inside due to deep gravitational potential well, leading to  trapped modes that survive for long durations.

 Chandra was awarded the Nobel prize for physics in 1983. 
His method of studying  diverse astrophysical topics involved applying  physical theories that had been corroborated experimentally, and then subjecting the 
 relevant  equations that followed from theory to rigorous and innovative mathematical analysis. No wonder that most of the new results he obtained were later confirmed by observations.  
 
\section { }
 
{\bf References}

 [1] Fowler, R. H., 1926, Mon. Not. R. Astr. Soc. 87, 114.
 
 [2] Chandrasekhar, S.,  Astrophys. J. 74, 81 (1931)
 
 [3] Anderson, W., Z. Phys. 56, 851 (1929)
 
 [4] Stoner, E. C. , Phil. Mag. 9, 944 (1930)
 
 [5] S. Chandrasekhar, An introduction to the study of stellar structure, University of Chicago 
 Press, Chicago (1939)
 
 [6] Landau, L., Phys. Z. Sowjetunion 1, 285 (1932) 

 [7]  Also see, Israel, W., in 300 years of Gravitation, eds. S. Hawking and W. Israel (Cambridge University
  Press, Cambridge, 1987); Srinivasan, G., J. Astrophys. Astr. 17, 53 (1996);    
   Blackman, E. G., arXiv: 1103.1342 [physics. hist-ph]

 [8] Baade, W. and Zwicky, F., Phys. Rev. 45, 138 (1934)

 [9] Baade, W. and Zwicky, F., Proc. Nat. Acad. Sci. 20, 254  (1934)

 [10] Bethe, H. A., Rev. Mod. Phys. 62, 801 (1990)

 [11] Demorest, P. B., Pennucci, T., Ransom, S. M., Roberts, M. S. E. and  Hessels, J. W. T., Nature 467, 1081 (2010)

 [12] Bhattacharya, D., Pramana 77, 29 (2011), and the references therein.

 [13] Zhang, W.,  Woosley, S. E. and  MacFadyen, A. I., Astrophys. J. 586, 356 (2003),  and the references therein.

 [14] Eddington, A. S., Observatory 58, 37 (1935)
 
 [15] A Raychaudhuri, Relativistic cosmology, I, Phys. Rev. 98, 1123 (1955)
 
 [16] Penrose, R., Phys. Rev. Lett. 14, 57 (1965)
 
 [17] Hawking, S. W. and Penrose, R., Proc. Roy. Soc. London A 314, 529 (1970)
 
 [18] Chandrasekhar, S., Astrophys. J. 97, 255 (1943). 
 
 [19] Chandrasekhar, S., Astrophys. J. 97, 263 (1943). 
 
 [20] Shu, F., Physical Universe (University Science Books, Mill Valley, California, 1982), for a simple explanation.
 
 [21] Binney, J. and Tremaine, S., Galactic Dynamics (Princeton University Press,
  Princeton, New Jersey, 1987)

 [22] David Merritt, arXiv: 1103.5446 [astro-ph.GA], for a recent discussion on 
 generalizing dynamical friction and stellar dynamics to relativistic regime. 
 
 [23] Bethe, H., Ζ. Phys. 57, 815 (1929)
 
 [24] Wildt, R., Astrophys. J. 89, 295 (1939) 
 
 [25] Wildt, R., Astrophys. J. 90, 611 (1939) 
 
 [26] Wildt, R., Astrophys. J. 93, 47  (1941)
 
 [27] Chandrasekhar, S., and Krogdahl, M. K., Astrophys. J. 98, 205 (1943)
 
 [28] Chandrasekhar, S., Astrophys. J. 100, 176 (1944) 
 
 [29] Chandrasekhar, S., Astrophys. J. 102, 223 (1945) 
 
 [30] Chandrasekhar, S., and Breen, F. H., Astrophys. J. 104, 430 (1946)
 
 [31] Chandrasekhar, S., Herzberg, G., Phys. Rev. 98, 1050 (1955)
 
 [32] Chandrasekhar, S., Astrophys. J. 128, 114 (1958)
 
 [33] Chandrasekhar, S., Radiative Transfer (Dover, New York, 1960)
 
 [34] Rau, A. R. P., J. Astrophys. Astr. 17, 113 (1996), and the references therein.
 
 [35] Moehs, D. P.,  Peters, J. and Sherman, J., IEEE Trans. Plasma. Sci.  33, 
  1786 (2005), and the references therein.
 
 [36] Holmes, A. J. T., Plasma Physics and Controlled Fusion 34, 653 (1992), and the references therein.
 
 [37] Cowling, T. G., Mon. Not. R. astron. Soc. 94, 39 (1934)
 
 [38] Backus, G. and Chandrasekhar, S., Proc. natn. Acad. Sci. 42, 105
  (1956)
 
 [39] Chandrasekhar, S., Astrophys. J. 124, 232 (1956)
 
 [40] Backus, G., Astrophys. J. 125, 500 (1957)
 
 [41] Chandrasekhar, S., Hydrodynamic and Hydromagnetic Stability (Oxford University
  Press, Oxford, 1961)
 
 [42] Frank, H. Shu, The Physics of Astrophysics, Vol. II (University Science Books, 
 Mill Valley, California, 1992)
 
 [43] Lynden-Bell, D. and Pringle, J. E., Mon. Not. R. Astr. Soc. 168, 603 (1974)

 [44] Balbus, S. and Hawley, J., Astrophys. J. 376, 214 (1991)
 
 [45] Parker, E.N., J. Astrophys. Astr. 17, 147 (1996) 
 
 [46] Chandrasekhar, S., Phys. Rev. Lett. 24, 611 (1970)
 
 [47] Friedman, J. L. and Schutz, B. F., Astrophys. J. 222, 281 (1978)
 
 [48] Everitt, C. W. F. et al., Phys. Rev. Lett. 106, 221101 (2011)
 
 [49] Andersson N.,  Astrophys. J. 502, 708 (1998)
 
 [50] Lindblom L., Owen B.J., Morsink S., 1998, Phys. Rev. Letters 80, 4843
 
 [51] Andersson N., Kokkotas K.D., Schutz B.F., Astrophys. J. 510, 2 (1999)
 
 [52] Owen B., Lindblom L., Cutler C., Schutz B.F., Vecchio A., Andersson
 N., Phys. Rev. D 58,  084020-1 (1998)

 [53] Pitkin, M.,  Reid, S.,  Rowan, S. and  Hough, J., Liv. Rev. Rel. 3, 
 3 (2000) (arXiv:1102.3355 [astro-ph.IM])

 [54] Chandrasekhar, S., The Mathematical Theory of Black holes (Clarendon Press, Oxford, 1983)
 
 [55] Hawking, S., Nature 248,  30 (1974)

 [56] Begelman, M. C.,  Blandford, R. D. and  Rees, M. J., Rev. Mod. Phys. 56, 255 (1984)
 
 [57] Schodel, R., Ott, T., Genzel, R., et al. 2002, Nature 419, 694
 
 [58] Vishveshwara, C.V.,  Nature 227, 936 (1970)

 [59] Press, W., Astrophys. J. 170, L105 (1971)
 
 [60] Chandrasekhar, S. and Friedman, J. L., Astrophys. J. 177, 745 (1972) 
 
 [61] Chandrasekhar, S.,  Proc. R. Soc. London A 343, 289 (1975)  
 
 [62] Chandrasekhar, S., and Detweiler, S., Proc. R. Soc. London A 344, 441  (1975)

 [63] Chandrasekhar, S., Proc. R. Soc. London A 349, 571 (1976)

 [64] Chandrasekhar, S. and Xanthopoulos, B. C.,  Proc. R. Soc. London A, 367, 1 (1979)

 [65] Chandrasekhar, S.,  Proc. R. Soc. London A 369, 425 (1980)

 [66] Chandrasekhar, S. and Xanthopoulos, B. C.,  Proc. R. Soc. London A, 378, 73 (1981)
 
 [67] Chandrasekhar, S.,  Proc. R. Soc. London A 392, 1 (1984)
 
 [68] Page, D. N., Phys. Rev. D 14, 1509 (1976)

 [69] Teukolsky, S.  A.,  Astrophys.  J. 185, 635 (1973).

 [70] Ferrari, V. and Gualtieri, L., Gen. Rel. Grav. 40, 945 (2008)

 [71] Penrose, R., Rev. Nuovo Cimento 1 (Special number), 252 (1969)

 [72] Zel$^\prime $dovich, Ya. B., JETP Lett. 14, 180 (1971)

 [73] Chandrasekhar, S., and Detweiler, S., Proc. R. Soc. London A 352, 325 (1977) 

 [74] Unruh, W., Phys.Rev.Lett. 31, 1265 (1973) 

 [75] Iyer, B. R. and Kumar, A., Phys. Rev. D 18, 4799 (1978)

 [76]  Chandrasekhar, S., in General Relativity: An Einstein Centenary Survey,
 eds. S. W. Hawking and W. Israel (Cambridge Univ. Press, New York, 1979)

 [77] Szekeres, P.,  Nature 228, 1183 (1970)

 [78] Khan, K. A. and Penrose, R., Nature 229, 185 (1971)

 [79] Chandrasekhar, S. and Ferrari, V.,  Proc. R. Soc. London A 396, 55 (1984)

 [80] Chandrasekhar, S. and Xanthopoulos, B. C.,  Proc. R. Soc. London A 408,  175 (1986)

 [81] Chandrasekhar, S. and Xanthopoulos, B. C.,  Proc. R. Soc. London A 410, 311 (1987)

 [82] Chandrasekhar, S., Selected Papers, Vol. 6 (The University of University Press, Chicago,
 1991)

 [83] Chandrasekhar, S., and Ferrari, V., Proc. R. Soc. London A 432, 247 (1991)

 [84] Chandrasekhar, S., and Ferrari, V., Proc. R. Soc. London A 434, 449 (1991)

 [85] Chandrasekhar, S., and Ferrari, V., Proc. R. Soc. London A 433, 423 (1991)

 \end{document}